\documentclass[12pt]{article}
\usepackage{graphicx}
\usepackage{amsmath}
\newcommand{\etc}{\textit{etc.}}
 \newcommand{\ie}{\textit{i.e.\ }}
 \newcommand{\eg}{\textit{e.g.\ }}
 
 \newcommand{\vs}{\textit{vs.}}

 \newcommand{\Eq}[1]{Eq.\ (\ref{#1})}

 \newcommand{\N}{\mathbf{N}}
 \newcommand{\I}{\mathbf{I}}
 \newcommand{\p}{\mathbf{p}}

\begin{document}
\title{Geometric construction of voting methods that protect voters' first choices}
\author{Alex Small}

\maketitle

\abstract{We consider the possibility of designing an election
method that eliminates the incentives for a voter to rank any other
candidate equal to or ahead of his or her sincere favorite.  We
refer to these methods as satisfying the ``Strong Favorite Betrayal
Criterion" (SFBC). Methods satisfying our strategic criteria can be
classified into four categories, according to their geometrical
properties.  We prove that two categories of methods are highly
restricted and closely related to positional methods (point systems)
that give equal points to a voter's first and second choices.  The
third category is tightly restricted, but if criteria are relaxed
slightly a variety of interesting methods can be identified.
Finally, we show that methods in the fourth category are largely
irrelevant to public elections. Interestingly, most of these methods
for satisfying the SFBC do so only ``weakly," in that these methods
make no meaningful distinction between the first and second place on
the ballot. However, when we relax our conditions and allow (but do
not require) equal rankings for first place, a wider range of voting
methods are possible, and these methods do indeed make meaningful
distinctions between first and second place.}

\tableofcontents

\section{Introduction}

\textbf{This is a draft.  I would appreciate feedback.  Public
discussion of this draft can be undertaken at
http://votingmath.blogspot.com.}

Voting theorists have known since the work of Gibbard and
Satterthwaite that voting systems using ranked ballots will give
incentives for insincere voting\cite{gibbard1973,Satterthwaite1975}.
Nonetheless, many people defend various voting reform proposals (\eg
Instant Runoff Voting) by claiming that their proposal will solve
the ``lesser of two evils" problem, and allow voters to support
their sincere favorite candidate \cite{jackson2004,anderson2006}. It
is easy to show that this claim is false for Instant Runoff Voting,
as there will still be cases in which voters have an incentive to
insincerely rank a ``lesser evil" in first place. Still, the claims
are common, suggesting that there is public interest in the design
of voting methods that eliminate the incentive to list a ``lesser
evil" in first place.  It is therefore worth exploring the extent to
which incentives for manipulation can be reduced, with particular
attention to incentives regarding first place rankings.  We can
analyze this by considering whether voting methods satisfy a
criterion called the ``Favorite Betrayal Criterion" (FBC), which
relates to the incentives voters face when deciding which candidate
to list in first place \cite{Venzke,smith}.  The FBC can be stated
as:

\newtheorem{definition}{Definition}
\begin{definition}
A voting method satisfies the \textbf{Favorite Betrayal Criterion}
(FBC) if there do not exist situations where a voter is only able to
obtain a more preferred outcome (\ie\ the election of a candidate
that he or she prefers to the current winner) by insincerely listing
another candidate ahead of his or her sincere favorite.
\end{definition}

A variety of methods are known to satisfy this statement of the
FBC\cite{Venzke,smith}.  One well-known example is Approval
Voting\cite{brams1983}.  However, most of these methods satisfy the
FBC by allowing a voter to rank one or more candidates equal to his
or her sincere favorite.  This is, in some sense, a ``weak" way of
protecting one's favorite candidate.  We will instead explore a
stronger version of the FBC, in which it is further stipulated that
a voter never has an incentive to rank another candidate \emph{equal
to} his or her favorite:

\begin{definition}
A voting method satisfies the \textbf{Strong Favorite Betrayal
Criterion} (SFBC) if there do not exist situations where a voter is
only able to obtain a more preferred outcome (\ie\ the election of a
candidate that he or she prefers to the current winner) by
insincerely listing another candidate ahead of or equal to his or
her sincere favorite.
\end{definition}

Note that neither definition requires that there never be situations
in which a more preferred outcome can be obtained by insincerely
voting another candidate ahead of one's favorite.  Such situations
can still exist with an FBC-compliant method.  However, the voter
should be able to do \textit{at least} as well by indicating some
other insincere ordering that leaves his or her favorite(s) at the
top of the list while insincerely changing the relative rankings of
the other candidates.

To make this concrete, consider a simple example of a method that
satisfies FBC and SFBC:  antiplurality voting.  In this method,
voters rank the candidates on their ballots, each candidate receives
one point for each ballot on which he or she is \emph{not} ranked
last, and the candidate with the most points wins. A voter has no
disincentive to rank his or her sincere favorite in first place, so
it complies with SFBC.  However, it is still a manipulable method,
in accordance with the Gibbard-Satterthwaite Theorem, since a voter
may have an incentive to insincerely rank some other candidate in
last place, if there is a close race between that candidate and the
voter's favorite.

Also, antiplurality voting illustrates our point about the
possibility of obtaining a more preferred outcome either by demoting
one's favorite OR by indicating some other ranking that leaves one's
favorite candidate(s) at the top of the list.  For instance, suppose
that in an antiplurality election with 4 candidates the top
contenders are a voter's second and third choices.  (Call these
respective candidates $c_2$ and $c_3$, for convenience.)  That voter
has a clear strategic incentive to insincerely rank $c_3$ in last
place.  Which candidate he or she ranks in first place on his or her
ballot is irrelevant, as all candidates except $c_3$ will receive
one point each.  In this case, ranking $c_2$ in first place will
have the same effect as ranking $c_2$ in second place and leaving
his or her sincere favorite (called $c_1$ for convenience) in first
place, as long as $c_3$ is ranked last.  Similar SFBC-compliant
methods can be designed in which the first and second choices each
receive one point, and candidates ranked lower receive specified
fractions of a point.

Note, however, that antiplurality voting only ``weakly" satisfies
the \emph{intent} of the SFBC, since the first place designation
given to the favorite is purely ceremonial.  There is no practical
distinction between ranking a candidate in first place or second
place.  Still, while there is no difference between first and second
place for  determining the outcome, there might be political or
social significance attached to the first place votes, \eg as a
measure of party strength.

The issue that we address here is whether other SFBC-compliant
methods exist, and the nature of such voting methods.  We will use
geometric techniques to show that SFBC-compliant methods can be
classified into 4 categories.  Two of the categories are highly
restricted and have some features similar to anti-plurality voting.
A third category is heavily restricted, but if the SFBC condition is
relaxed to FBC it includes a variety of recently-proposed methods
\cite{Venzke,smith}.  We will show that the fourth category is
largely irrelevant to public elections.

Finally, before beginning our formal analysis, a note on voter
incentives:  One could argue that our analysis is irrelevant because
individuals have almost no incentive to vote insincerely if
elections decided by a single vote are rare. A trivial response is
that \textit{almost} no incentive is not the same as \textit{no}
incentive. Our more serious response is that in practice one can
consider strategic choices facing campaigners and activists seeking
to influence voters.  Should they advise a faction of voters with
similar preferences to vote sincerely or strategically? If enough
voters heed the advice of a campaigner, especially when using a
complicated ranked method that makes strategic incentives opaque to
the non-expert, then the decisions of an individual performing
strategic calculations can indeed influence the course of an
election.

%(For an example of campaigners advising groups of
%like-minded voters, consider ``above the line" voting in Australia,
%where many voters choose to cast a ballot marked with a ranking
%decided by party strategists, rather than filling out their own
%ranking of all of the candidates listed.)

\section{Ballot Design}\label{sec:ballotdesign}

Voting theorists have proposed a wide range of election methods
using a wide range of ballot types.  Here we restrict our attention
to methods which can be conducted with ballots on which each voter
assings a rank to each candidate, and no other information is
indicated on the ballot for the purpose of determining the election
outcome.  We make no assumptions about whether more than one
candidate can be assigned the same rank, or whether a voter can
leave some ranks unused.  For instance, one could imagine a method
in which a candidate receives 5 points if assigned the first rank, 4
points if assigned the second rank, 3 points if assigned the third
rank, and so on down to the sixth rank (for which a candidate
receives zero points).  In this method, a voter might assign a
favorite candidate the first rank, a compromise candidate the second
rank, and all remaining candidates in the last rank (or not list the
remaining candidates, formally equivalent to listing them in the
sixth rank).  A rank has still been assigned to every candidate,
even if some ranks were not used and other ranks were used multiple
times.

Likewise, we do not exclude from consideration methods in which the
number of ranks is less than the number of candidates.  For
instance, in Approval Voting every candidate is either approved or
disapproved (equivalent to ranking every candidate in either the
first or the second slot on a ballot) and the winner is the
candidate ranked ``approved" on the greatest number of
ballots\cite{brams1983}.  As another example, in the ``plurality" or
``first past the post" elections commonly used in the United States,
only one candidate receives a vote, implicitly equivalent to a first
rank, and all of the other candidates are implicitly ranked in the
second position.

\section{Criteria}

For the sake of simplicity, we impose several criteria on the
election methods under consideration.

\begin{enumerate}
\item Anonymity:  The outcome doesn't depend upon which voter
submits which ballot, but only on how many voters cast ballots of
each type. Formally, the outcome is function of $\{n_k\}$, where
$n_k$ is the number of voters submitting ballots that indicate a
given preference order $P_k$.

\item Neutrality:  All candidates are treated equally, so that
the level of support required for victory is the same for all
candidates.  Formally, if every voter changes his or her ballot in
such a way that the winning candidate $c_1$ is swapped with a
candidate $c_2$ (while all other preferences are left unchanged),
then $c_2$ should win.  Conversely, if every voter changes his or
her ballot in such a way that the winning candidate $c_1$ retains
the same ranking but losing candidates $c_2$ and $c_3$ are swapped
(while all other preferences are left unchanged), then $c_1$ should
still win.

\item No Turnout Quota:  The number of voters participating in an
election is irrelevant to the result.  Only the \emph{fraction} of
the electorate casting ballots of a particular type is relevant.

\item Linearity:  The conditions for a candidate to win can be
expressed by a series simple inequalities that are linear in the
tallies of the ballot types.  Formally, the inequalities that must
be satisfied for candidate $c_i$ to win involve conditions of the
form:
    \begin{equation}
        \sum_{k} u_{ijk} \cdot n_k > 0
        \label{eq:inequalities}
    \end{equation}
where $\{u_{ijk}\}$ are constant coefficients (some of which may be
negative), $i$ refers to the candidate, $j$ indexes the condition
being checked, and $k$ indexes the preference orders.

Note that because of the No Turnout Quota criterion the right hand
sides of the inequalities can be set to zero without loss of
generality:   Suppose the right hand side of an inequality were some
constant $\alpha$.  Because the numbers of voters with each
preference order will sum to $n_V$ (the total number of voters), we
can write the righthand side as $\alpha \cdot
\left(\sum_{k}n_k\right)/n_V$, and then rewrite the inequality as
$\sum_{k} u_{ijk} \cdot n_k - \alpha \cdot n_k/n_V = \sum_{k} \left(
u_{ijk} - \alpha/n_V \right) \cdot n_k > 0$. With a suitable
redefinition of the coefficients, we can write the inequality with a
zero on the righthand side.

\item Decisiveness:  There is always a single winning candidate,
except in the case of ties.  Formally, a necessary condition for a
tie is that at least one of the expressions being evaluated (\ie an
expression of the form $\sum_{k} u_{ijk} \cdot n_{k}$) to determine
the outcome is equal to zero.

\end{enumerate}

Our Anonymity and Neutrality criteria are violated in some notable
cases.  For instance, American Presidential elections violate our
formulation of the Anonymity criterion, since additional supporters
only help a candidate if those supporters live in states that the
candidate needs to win.  Also, a recall election might be considered
in violation of Neutrality if, for instance, an incumbent with
49.9\% support is removed for lacking a majority, and succeeded by
somebody who wins 35\% support in a 3-way race. Nonetheless, the
Anonymity and Neutrality criteria are satisfied in most public
elections held in democratic societies.  We impose them in this work
to gain considerable simplicity without undue loss of generality.

The Linearity criterion, while not always explicitly discussed, is
satisfied in every seriously proposed election method that we are
aware of. Still, one could invent hypothetical election methods
utilizing inequalities that are nonlinear in ballot tallies, so we
must impose Linearity explicitly. Fortunately, it is a very weak
criterion:  In our analysis we will examine criteria satisfied by
boundaries and their normal vectors.  We expect that if one wanted
to analyze nonlinear methods, most of the results would carry over
because they impose constraints on the orientation of boundaries.

%If a nonlinear method used inequalities that gave rise to smooth
%boundaries (\ie normal vectors exist and vary as continuous functions
%except for discontinuities where boundaries meet) the same results
%would hold.

The No Turnout Quota criterion eliminates from our analysis any
election rules in which the result depends on the absolute number of
ballots cast, \eg elections in which a quorum of eligible citizens
must support a result for it to be valid.  In such situations,
abstention may become a viable strategy for voters, a complication
that we neglect here.  Due to our Decisiveness criterion, the only
elections without a valid winner will be elections that end in ties,
and the set of tied elections will be of lower dimension than the
set of possible elections (since a linear expression must be exactly
equal to zero for a tie to occur).

\section{Outline}

The rest of this paper proceeds as follows:  In the next section we
will outline the geometrical formalism that we use to describe
election methods. Our formalism is closely related to that used by
Saari\cite{saari1995}, representing the set of ballots cast as a
point in the unit simplex.  In this formalism, an election method
partitions the simplex into regions, each region corresponding to
victory by a particular candidate.  The components of the normal
vectors to the boundaries are related to the coefficients
$\{u_{ijk}\}$ introduced in our definition of the Linearity
criterion.  Using reasoning similar to Saari's proof of the
Gibbard-Satterthwaite Theorem, we will show that the boundaries
between these regions determine the strategic incentives confronting
a voter.  We will then derive requirements that each boundary (and
its normal vector) must satisfy in order for a method to comply with
the SFBC.

Based on the geometric properties of the boundaries, we will
classify SFBC-compliant election methods into 4 distinct categories.
The most important result in this work concerns methods in which
each boundary satisfies a different.  These methods correspond to
certain positional methods (elections in which a candidate receive
points according to the number of voters assigning him each rank).
In the second category, some of the boundaries satisfy multiple
conditions, and we will prove that methods in this category can be
thought of as a hybrid of runoffs and positional methods, akin to
the Bucklin method\cite{Tideman2006}.  In the third category, all of
the boundaries satisfy multiple conditions, and we will give
examples of methods in this category.  Finally, in the fourth
category every boundary satisfies every possible condition, and we
will argue that methods in this category are uninteresting for
public elections.

\section{Formalism}

\subsection{Vector Notation for Electorates}
Because of our Anonymity criterion, the outcome of the election can
be determined if we know the number of voters indicating each
preference order $P_i$.  In addition, as discussed above, our No
Voter Turnout Quota criterion implies that the total number of
voters $n_V$ participating is irrelevant to the election outcome. It
then follows that a listing of the \emph{fraction} of voters
indicating each preference order $P$ contains sufficient information
to specify the election outcome.  We will therefore follow the lead
of Saari and describe the electorate with what Saari refers to as a
normalized profile vector $\p$, where each component $p_k = n_k/n_V$
indicates the fraction of the voters who cast a ballot indicating a
preference $P_k$.  (This definition for the components also defines
the basis that we will use for working in this vector space.)

The dimensionality $d$ of the space that these vectors reside in
depends on the number of candidates $n_c$ and whether we allow
voters to cast ballots where some candidates are ranked equal.  In
the case where equal rankings are disallowed (recall that the Strong
FBC rules out the need for ballots listing multiple candidates in
first place), the dimensionality of the space is $n_c!$.

In any case, profile vectors will lie in the unit simplex $Si(d)$,
defined as the set of all vectors in $d$-dimensional space such that
all of their components (in our chosen basis) are non-negative and
add up to unity.  It is convenient to express this summation of
components to unity as a condition on the inner product
$(\mathbf{p}, \I) =  \sum_k p_k \cdot 1$:
\begin{equation}
    (\p, \I) = 1
    \label{eq:unity}
\end{equation}
where $\I$ is a vector for which all of the components are 1.  The
vector $\I$ is significant in this work in part because the relation
$\left(\mathbf{p},\I\right) = 1$ enables us to simplify certain
results, and also because $\I/d$ lies at a symmetry point (the
center of the simplex).

We can also formulate our victory conditions in terms of inner
products, and re-write \Eq{eq:inequalities} as:
\begin{equation}
    \sum_{k} u_{ijk} \cdot n_k = \sum_k u_{ijk} \cdot p_k \cdot n_V = (\p,\mathbf{u_{ij}}) \cdot  n_V> 0
    \label{eq:inequalities2}
\end{equation}
where $\mathbf{u_{ij}}$ is a vector of coefficients defining the
$j$th condition for candidate $c_i$ to win.  Because $n_V>0$,
\Eq{eq:inequalities2} can just as easily be written as $(\p,
\mathbf{u_{ij}})>0$.

\subsection{Boundaries}

An election method can be considered as a procedure for dividing the
simplex into $n_c$ regions, each region corresponding to the
election of one of the $n_c$ candidates.  Geometrically, we can say
that an election method specifies the boundaries between regions and
assigns a winner to each of the $n_c$ regions.  In this view, the
Neutrality criterion specifies the symmetry of the boundaries.

Consider the boundary between a region where candidate $c_i$ wins
and a region where candidate $c_j$ wins.  Mathematically, this
boundary can be specified piecewise in terms of its normal vector
$\N_{ij}$.  The vector $\N_{ij}$ points outward from the boundary
and into the region where candidate $c_i$ wins, while $-\N_{ij}$
points outward from the boundary and into the region where $c_j$
wins.  A basic fact of analytical geometry is that all of the points
in a flat surface will have the same projection onto the normal
vector.  In our formalism, this means that $(\p,\N_{ij})$ is
constant along the boundary.  We can set the constant to whatever
value we like and still have a geometrically valid definition of a
boundary.  However, in keeping with \Eq{eq:inequalities2} we will
assume that the inner product is zero, along a boundary, \ie\:
\begin{equation}
    \left(\mathbf{p},\N_{ij}\right) = 0
    \label{eq:victory}
\end{equation}
The choice of zero for the right hand side corresponds to the notion
that boundaries represent tied results, in which case two weighted
ballot counts exactly cancel.  Also, there is no loss of generality
here, due to our assumption that $(\p,\I) = 1$.  Suppose that one
wanted to define boundaries so that $(\p,\N_{ij})$ is equal to some
constant value $a$.  We can rewrite our new victory condition
$(\p,\N_{ij}) = a$ as $(\p,\N_{ij}) = (\p,a\I)$, or
$(\p,\N_{ij}-a\I) = 0$, which amounts to a redefinition of the
normal vector in \Eq{eq:victory}.

Note that because the boundaries may be defined piecewise (giving
rise to kinks or folds in profile space), satisfaction of
\Eq{eq:victory} may not be a \emph{sufficient} condition for a
profile to lie on the $i-j$ boundary.  The piecewise specification
of the boundaries is the subject of the next section.

The Neutrality condition imposes certain requirements on the
boundaries.  For instance, suppose a profile vector $\p$ lies on the
$i-j$ boundary defined by the normal vector $\mathbf{v}$, \ie\ $(\p,
\mathbf{v}) = 0$ and $\mathbf{v}$ points from the boundary into the
region where $c_i$ wins.  If every voter were to swap candidates
$c_i$ and $c_j$ on his or her ballot, then the profile vector should
still be somewhere on the $i-j$ boundary (although perhaps not on a
portion with normal vector $\mathbf{v}$, if the boundary is defined
piecewise).  Voters swapping candidates $c_i$ and $c_j$ on their
ballots corresponds to swapping components of the profile vector
$\p$.  This is a linear operation, and so its effect on the profile
vector $\p$ can be represented by a matrix $S_{i,j}$ (which we will
call a ''symmetry operator") acting on $\p$.  There should hence be
some other normal vector $\mathbf{v'}$ (possibly equal to
$-\mathbf{v}$) that defines some  portion of the (possibly
piecewise-defined) $i-j$ boundary such that $(S_{i,j}\p,
\mathbf{v'}) = 0$.

It is easy to show that the operator $S_{i,j}$ is unitary or
self-adjoint (in the language of linear algebra), and this enables
us to relate the inner normal vectors $\mathbf{v'}$ and
$\mathbf{v}$.  The fact that $S_{ij}$ is unitary implies that:
\begin{equation}
(S_{ij}\p, \mathbf{v'})=(\p,S_{i,j}\mathbf{v'}) = 0
 \label{eq:unitary}
\end{equation}

The condition $(\p, \mathbf{v}) = 0$ therefore implies that $(\p,
S_{i,j}\mathbf{v'}) = 0$, and hence $\mathbf{v}$ and
$S_{i,j}\mathbf{v'}$ must be linearly dependent, \ie\ they must be
proportional to each other by some constant scalar factor.  The
magnitude of the factor is irrelevant for our purposes, and whether
it is positive or negative depends on how we define the orientations
of $\mathbf{v}$ and $\mathbf{v'}$ relative to the $i-j$ boundary.
However, in the special case where $\mathbf{v}$ and $\mathbf{v'}$
are also linearly dependent (\ie\ the electorate obtained by
swapping $c_i$ and $c_j$ on every ballot lies on the same portion of
the piecewise defined boundary) it must be the case that
$\mathbf{v'} = -\mathbf{v}$.  In that special case, one of the
vectors ($\mathbf{v}$ or $\mathbf{v'}$ defines the inner normal
$\N_{ij}$ pointing toward the region where $c_i$ wins, and the other
vector defines the inner normal $\N_{ji}$ pointing from the same
boundary toward the region where $c_j$ wins.

A similar line of reasoning shows that if a vector $\mathbf{v}$ is
normal to some portion of the $i-j$ boundary, then
$S_{j,k}\mathbf{v}$ should be normal to some portion of the $i-k$
boundary.  We thus see that as a consequence of the Neutrality
criterion the boundaries between victory regions are related by
symmetry operations involving the exchange of candidate names.

\subsection{Election Methods as Sequential Procedures}

Having established the vector notation for defining election methods
according to the vectors normal to boundaries between victory
regions, let us now consider how boundaries might be defined
piecewise.  Election methods can generally be specified as procedure
defined by a series of if-then-else statements regarding the numbers
of voters submitting different ballot types.  For instance, the
simple (and SFBC-compliant) method of antiplurality voting could be
formally specified as:

\begin{tabular}{lp{3in}}
IF & The number of voters listing $c_1$ in last place is less than the number of voters listing $c_2$ in last place, \\
AND & The number of voters listing $c_1$ in last place is less than the number of voters listing $c_3$ in last place, \\
AND & ... \\
THEN & Candidate $c_1$ wins.\\
ELSE\\
IF & The number of voters listing $c_2$ in last place is less than the number of voters listing $c_1$ in last place, \\
AND & The number of voters listing $c_2$ in last place is less than the number of voters listing $c_3$ in last place, \\
AND & ...\\
THEN & Candidate $c_2$ wins.\\
ELSE\\
etc.
\end{tabular}
\noindent If the conditions are expressed as sets of linear
inequalities, we get:

\begin{tabular}{lp{3in}}
IF & $(\mathbf{\p, u_{11}}) > 0$ \\
AND & $(\mathbf{\p, u_{12}}) > 0$ \\
AND & ...\\
THEN & Candidate $c_1$ wins.\\
ELSE\\
IF & $(\mathbf{\p, u_{21}}) > 0$ \\
AND & $(\mathbf{\p, u_{32}}) > 0$ \\
AND & ...\\
THEN & Candidate $c_2$ wins.\\
ELSE \\
etc.
\end{tabular}

\noindent where the vectors $\{\mathbf{u_{ij}}\}$ are chosen so that
$(\p, \mathbf{u_{11}})$ is (for this particular example) the
difference between the number of last place votes received by
candidate $c_2$ and candidate $c_1$, and so forth.

Notice that in the text description of the voting method the
conditions for candidate $c_2$ to win could be obtained by swapping
the labels 1 and 2 for the candidates, in accordance with our
Neutrality criterion.  If the conditions are translated into sets of
linear inequalities, then (as discussed above) the vectors
$\mathbf{u_{ij}}$ defining those conditions are related by the
symmetry operator $S_{1,2}$.  In general, once we define a set of
sufficient conditions under which some candidate $c_i$ wins, we can
invoke Neutrality to define conditions for any other candidate to
win.

We will refer to a set of conditions related by symmetry operations
as a ''stage" of an election method.
\begin{definition}
Given a set of linear inequalities such that the simultaneous
satisfaction of those inequalities is a sufficient for a particular
candidate $c_i$ to win (and the vectors used to define the
associated linear inequalities), we can generate a \textbf{stage} of
victory conditions by applying swap operators $S_{j,k}$ to those
vectors for \textit{all} $j$ and $k$.  Cases where $i=j$ or $i=k$
will produce conditions for some other candidate to win.  Cases
where $j \neq i$ and $k \neq i$ will produce alternative conditions
for $c_i$ to win.
\end{definition}

The popular Instant Runoff method is an example of a method where a
single stage will have multiple conditions for the same candidate to
win.  In Instant Runoff, a candidate can win if he or she survives
successive eliminations to become one of the 2 final candidates
considered, and receives majority support over the other finalist.
For that method, there would be multiple scenarios under which the
candidate could win, corresponding to different opponents in the
final round.

Whether a stage of conditions has $n_c$ conditions (1 per candidate)
or some multiple of $n_c$ (\ie\ multiple ways for each candidate to
win) depends on the vectors defining the primary set of inequalities
used to generate the stage.  If we have a series of inequalities
defining a sufficient set of conditions for $c_i$ to win, and if
applying \textit{any} swap operator $S_{j,k}$ ($j,\ k \neq i$) to
\textit{any} of the vectors associated with that set of conditions
generates another vector of that set of conditions, then the only
swap operation that changes the set of inequalities is one that
changes the name of the winner, and so there are only $n_c$
conditions in that stage (one per candidate).  Otherwise, the number
of conditions in that stage will be a multiple of $n_c$.  This fact
will prove to be useful later.

The sets of conditions for different candidates to win in a stage
must be mutually exclusive, so that a stage produces a unique
winner. However, the conditions need not be exhaustive, \ie\ a stage
could select no winner.  Numerous election methods can be expressed
as procedures with stages that sometimes yield no winner.  For
instance, if a method elects a candidate listed in first place by a
majority of the voters, and uses some alternative procedure when
there is no majority favorite, that method will be one in which the
first stage does not always yield a winner. When that is the case,
our Decisiveness criterion requires that the first stage of the
method be augmented by a series of subsequent stages so that the
method yields a unique winner in every case except ties, which will
correspond to a lower-dimensional subset of the unit simplex that
the electorate vectors reside in.

Also, although a stage of victory conditions can be generated by
applying swap operations to a single set of conditions for a
particular candidate to win, it may be that not all of the vectors
defining that primary set of conditions are necessary to specify the
election method.  Some of the vectors in that primary set used to
generate the stage may also be related to each other by swap
operations.  For instance, if a set of conditions for a candidate to
win consists of 5 linear inequalities, it may be that the vectors
defining some of those inequalities can be derived from each other
by swap operations.  It will be useful later on in this work to
consider the minimal number of vectors needed to specify all of the
inequalities in a stage of an election method.  We therefore
introduce this definition:
\begin{definition}
A \textbf{Minimal Set of Generators} $M$ for a stage $s$ of an
election method is a set of vectors $\{ \mathbf{v}_i \}$ such that:
\begin{itemize}
    \item Every vector defining every inequality in that stage can be obtained from a vector
    in $M$ by a suitable sequence of swap
    operations.

    \item No vector in $M$ can be obtained from any other vector in $M$ by any sequence of swap
    operations.
\end{itemize}
The number of vectors in $M$ is said to be the number of generators
for $s$.

\end{definition}
Note that the minimal set $M$ for a given stage $s$ is not unique.
If the same swap operator were applied to every vector in $M$, we
would still have another minimal set of generators for the same
stage $s$.

Finally, it is worth noting that when boundaries are specified
piecewise the same vector may define different boundaries under
different circumstances.  Consider, for instance, 3 candidates in an
election conducted with Instant Runoff.  Suppose that no candidate
is the first choice of a majority but candidate $c_1$ is the first
choice of the greatest number of voters.  Suppose also that
candidate $c_1$ can defeat candidate $c_2$ in a one-to-one contest,
but he or she cannot defeat $c_3$.  A necessary condition in
determining whether $c_1$ or $c_3$ wins is whether $c_2$ has more
first place votes than $c_3$ or fewer first place votes than $c_3$,
and so the vector $\mathbf{v}$ used to specify that linear
inequality is the vector normal to a portion of the $1-3$ boundary.
However, if $c_1$ could defeat $c_3$ in a one-to-one contest but
could not defeat $c_2$, then that same vector $\mathbf{v}$ would be
normal to a portion of the $1-2$ boundary.  We therefore see that
some of the vectors defining our linear inequalities may be normal
to multiple boundaries, depending on the circumstances.

\section{Voter Incentives and Geometry}

Whether or not a voter has an incentive to vote insincerely depends
on the boundaries between the victory regions for different
candidates.  Suppose, for instance, that the condition for candidate
$c_i$ to win instead of $c_j$ is $(\p,\N_{ij})>0$ (with the case
$(\p,\N_{ij})>0$ corresponding to a victory for candidate $c_j$).
The inner product $(\p,\N_{ij})$ is in essence a weighted sum over
the ballot types of the number of each type of ballot submitted,
with the ballot numbers corresponding to the components of the
vector $\p$ and the weighting factors corresponding to the
components of the normal vector $\N_{ij}$.  A voter therefore has an
incentive to submit whichever ballot type will receive the greatest
weight (positive or negative, depending on their preference). Voters
who prefer $c_i$ to $c_j$ will want to cast a ballot of a type
corresponding to the largest positive component of $\N_{ij}$, and
voters who prefer $c_j$ to $c_i$ will want to cast a ballot of a
type corresponding to the largest negative component of $\N_{ij}$.

This line of reasoning was used by Saari in his proof of the
Gibbard-Satterthwaite Theorem \cite{saari1995}, where he showed that
the theorem requires all of the components of the normal vectors to
be either the same positive number of the same negative number.  He
then went on to show that election methods corresponding to
boundaries defined by such normal vectors inevitably lead to
paradoxes.  Here, we will show that when this line of reasoning is
applied to the SFBC we can obtain somewhat less stringent conditions
on the normal vectors.

\subsection{\label{sec:vectors} SFBC-Compliant Methods}

Consider a voter whose sincere favorite candidate is $c_1$, and who
prefers some candidate $c_i$ ($i$ may or may not be equal to 1) to
some other candidate $c_j$ ($j\neq 1$).  If the election is a close
race between $c_i$ and $c_j$, then the condition determining the
election outcome will be whether $(\p,\N_{ij})$ is positive ($c_i$
wins) or negative ($c_j$ wins). As argued above, this voter will
want to cast a ballot of a type that corresponds to the largest
component(s) of $N_{ij}$.  In order for the method to comply with
the SFBC, that ballot type must list $c_1$ in first place.  A
similar analysis holds for any other voter who prefers $c_i$ to
$c_j$, and so the vector $\N_{ij}$ must have at least $n_c-1$
elements with the same maximum (largest positive) value.  Of those
maximum elements, \textit{at least} one must correspond to a
preference order listing each of the candidates other than $c_j$ in
first place.  Otherwise, there will be voters with an incentive to
list some candidate other than their favorite in first place.

A similar analysis shows that the normal vector $\N_{ij}$ must have
\textit{at least} $n_c-1$ minimum (largest negative) elements, and
of those at least one must correspond to a preference order listing
each of the candidates other than $c_i$ in first place.  This leads
us to our first significant result:
\newtheorem{theorem}{Theorem}
\begin{theorem} \label{thm:FBC}
    If a voting method complies with the SFBC, then any normal
    vector being used to define the $i-j$ boundary must satisfy the following conditions:
    \begin{enumerate}
        \item  At least $n_c-1$ components must have the same
        maximum (largest positive) value, and for each candidate
        other than $c_j$ there must be at least one component of
        $\N_{ij}$ corresponding to a preference order with that
        candidate in first place.

        \item  At least $n_c-1$ components must have the same
        minimum (largest negative) value, and for each
        candidate other than $c_i$ there must be at least one
        component of $\N_{ij}$ corresponding to a preference
        order with that candidate in first place.
    \end{enumerate}
\end{theorem}

Also, our Neutrality condition implies that the normals to different
boundaries can be obtained by the application of swap operators,
which would exchange candidates in preference orders and
correspondingly swap components of the normals.  Moreover, if a
vector satisfies the conditions for $\N_{ij}$ in an SFBC-compliant
method, then multiplying it by $-1$ exchanges the maximum (largest
positive) and minimum (largest negative) components.  Because the
conditions for normals are stated in terms of minimum and maximum
components, it follows that multiplying a suitable $\N_{ij}$ vector
by $-1$ gives a suitable $\N_{ji}$ vector.

Note that the requirements laid out in Theorem \ref{thm:FBC} are
necessary but not sufficient for a method to satisfy SFBC. It is not
enough for a method to define victory conditions in terms of vectors
satisfying the requirements of the theorem.  The conditions must
also be arranged in such an order that non-satisfaction of a
condition only leads to certain results.  When an election result
changes because an inequality of the form $(\p, \mathbf{v})>0$ is no
longer satisfied, and when the vector $\mathbf{v}$ satisfies the
requirements for the normal $\N_{12}$, the only possible outcome
should be that the election result is now the victory of candidate
$c_2$ instead of $c_1$.

As was remarked above, when boundaries are defined piecewise, the
same vector may define multiple boundaries, depending on the
situation.  This does not necessarily pose a problem for the normal
vectors specifying a SFBC-compliant method.  A normal vector
$\mathbf{v}$ could satisfy the necessary conditions for the normal
vectors $\N_{ij}$ and $\N_{ik}$ ($j \neq k$).  In that case, for any
candidate other than $c_i$ there must be at least one element of
$\mathbf{v}$ that has the minimum (largest negative) value and
corresponds to a preference order with that candidate in first
place. Also, for any candidate other than $c_j$ there must be at
least one element of $\mathbf{v}$ that has the maximum (largest
positive) value and corresponds to a preference order with that
candidate in first place, and for any candidate other than $c_k$
there must be at least one element of $\mathbf{v}$ that has the
maximum (positive) value and corresponds to a preference order with
that candidate in first place. The second pair of conditions imply
that for \textit{every} candidate there must be at least one element
of $\mathbf{v}$ with the maximum (largest positive) value and
corresponding to a preference order with that candidate in first
place.  This implies that $\mathbf{v}$ satisfies the conditions for
\textit{any} normal vector $\N_{ix}$ ($x \neq i$) in an
SFBC-compliant method.  This leads to the following useful result:
\begin{theorem}\label{thm:normal1}
If a vector $\mathbf{v}$ satisfies the necessary conditions for the
normal vectors $\N_{ij}$ and $\N_{ik}$ (assuming $j \neq k$) in an
SFBC-compliant method, $\mathbf{v}$ satisfies the necessary
conditions for \textit{any} normal vector $\N_{ix}$ ($x \neq i$) in
an SFBC-compliant method.
\end{theorem}

Analogous reasoning leads to the following result:
\begin{theorem}\label{thm:normal2}
If a vector $\mathbf{v}$ satisfies the necessary conditions for the
normal vectors $\N_{ji}$ and $\N_{ki}$ ($j \neq k$) in an
SFBC-compliant method, $\mathbf{v}$ satisfies the necessary
conditions for \textit{any} normal vector $\N_{xi}$ ($x \neq i$) in
an SFBC-compliant method.
\end{theorem}

Also, the same considerations lead to one more result concerning
normal vectors satisfying multiple conditions:
\begin{theorem}\label{thm:normal3}
If a vector $\mathbf{v}$ satisfies the necessary conditions for the
normal vectors $\N_{ij}$ and $\N_{kl}$ ($i \neq k$ and $j \neq l$)
in an SFBC-compliant method, $\mathbf{v}$ satisfies the necessary
conditions for \textit{any} normal vector $\N_{xi}$ (for all
admissible values of $x$ and $y$) in an SFBC-compliant method.
\end{theorem}

We therefore see that a vector normal to a boundary in an
SFBC-compliant method can fall into 3 categories, defined as
follows:
\begin{definition}
We will refer to 3 different types of normal vectors for the
boundaries between victory regions.
\begin{enumerate}
 \item A \textbf{Type 1 vector} \textit{only} satisfies the requirements necessary for the boundary between 2 particular victory regions.
 \item A \textbf{Type 2 vector} satisfies the requirements for \emph{all} of the boundaries to the victory region for a single candidate $c_i$.
 \item A \textbf{Type 3 vector} satisfies the requirements for all of the boundaries between all of the victory regions defined by this election method.
\end{enumerate}
\end{definition}

Note that while we have given these results in terms of compliance
with SFBC, the theorems given here also apply to FBC-compliant
methods if weakened slightly to include the possibility of ranking
two candidates together in first place.

\subsection{Geometric Classification of SFBC-Compliant Methods}

Having considered the requirements that individual normal vectors
must satisfy for an election method to satisfy the SFBC, let us now
consider how we can use these requirements to classify the stages
that comprise an election procedure.  Each stage will consist of a
series of linear inequalities defined by vectors that are in turn
related to each other by swap operators.  Each vector may satisfy
the requirements for a single boundary (case 1 above), multiple
boundaries on the same region (case 2 above), or all possible
boundaries (case 3 above).  Given those possibilities, the following
taxonomy will be useful for the results that follow:
\begin{definition}
The stages of an election method compliant with the SFBC shall be
referred to as coming in 4 different types.
\begin{enumerate}
\item[1]  A stage of an SFBC-compliant election method will be referred to as a \textbf{Type 1 stage}
if each vector defining a victory condition is of Category 1, \ie\
it only satisfies the requirements for a single boundary in an
SFBC-compliant method.  So, if a vector defining a condition for
candidate $c_i$ to win satisfies the requirements for the normal
vector $\N_{ij}$ it will not satisfy the requirements for the normal
vector $\N_{ik}$ ($k \neq j$).

\item[1b]  A stage of an SFBC-compliant election method will be referred to as a \textbf{Type 1b stage}
if some of the vectors satisfy a single condition (as in Type 1
stages) and other vectors are of category 2, \ie\ they satisfy
requirements for multiple boundaries of the same region (\ie\ some
vectors are of Category 2, \ie\ they satisfy the requirements for
$\N_{ij}$ and $\N_{ik}$) in an SFBC-compliant method.

\item[2]  A stage of an SFBC-compliant election method will be referred
to as a \textbf{Type 2 stage} if \textit{all} of the vectors
defining a condition for a candidate to win are of Category 2, \ie\
they satisfy the requirements for multiple boundaries of the same
region. For these types of stages, any vector defining a condition
for candidate $c_i$ to win will satisfy the requirements for
$\N_{ij}$ and $\N_{ik}$ (for all $k \neq j$).

\item[3]  A stage of an SFBC-compliant election method will be referred to as a
\textbf{Type 3 stage} if at least one of the vectors defining a
victory condition is of Category 3, \ie\ it satisfies the necessary
requirements for all possible boundaries.

\end{enumerate}
\end{definition}

This taxonomy is somewhat arbitrary, but the types are
non-overlapping and are related to key results proved below.  In
what follows we will prove that Type 1 stages are heavily restricted
and are related to the ''positional methods" that Saari has studied
extensively \cite{saari1995}.  We will then prove that Type 1b
stages are also heavily restricted, and are again related to
positional methods.  Type 2 stages are somewhat more varied and
cannot be easily categorized, but we will prove that they are
nonetheless subject to significant restrictions.  We will also show
that Type 2 stages occur in methods that have been studied by
others, if SFBC is relaxed to FBC \cite{Venzke,smith}.

We will not devote much attention to Type 3 stages, as it is
difficult to conceive of a socially desirable election method that
makes use of Type 3 stages.  We are not aware of any methods that
use Type 3 stages, and for the case of 3 candidates Type 3 stages
turn out to be rather tricky to construct in a manner that avoids
multiple winners (\ie\ it is tricky to construct a Type 3 stage in
which the conditions for different candidates to win are mutually
exclusive).  The reason for these complications in the case of 3
candidates is that if there are 3 candidates and $3!=6$ preference
orders (\ie\ 6 components to a normal vector) then 3 of those
components must have the same maximum value and 3 must have the same
minimum value.  This tightly constrains the number of possible
orientations for the vectors, and so it turns out to be necessary to
construct several conditions of the form $g_1 \leq (\p,\mathbf{v})
\leq g_2$ (where $g_1$ and $g_2$ are constants) to ensure that the
victory regions thus defined are non-overlapping.  When these
conditions are constructed and examined, it is difficult to
interpret them in terms of considerations usually used in the
construction of election method (\eg\ majority support, point
totals, or being one of the top 2 candidates to make a runoff).

The constraints are less significant in the case of 4 or more
candidates, but generally election method designers prefer methods
that can be stated in terms of simple criteria that can be just as
easily described regardless of the number of candidates.  If the
conditions defining an election method cease to make sense when the
number of candidates is reduced to 3, the election method is
unlikely to be of interest.  Moreover, even ignoring the issue of
the number of candidates, in elections using Type 3 stages it is
possible that two voters with the same favorite candidate may have
completely opposite effects on that candidate's election prospects,
because of the way that vectors in Type 3 methods have both a
maximum and a minimum component corresponding to preferences with
the same candidate in first place.  While it is often the case that
casting a ballot with a particular preference order (even a
preference order listing one's sincere favorite in first place) will
be sub-optimal (hence the interest in analyzing strategic
incentives), methods with Type 3 stages are more pathological than
most other methods (including methods that fail to satisfy FBC or
SFBC).  For this reason, we will not analyze Type 3 stages in any
detail here, and the remainder of this work will focus on the other
Types of SFBC-compliant stages.

\section{Type 1 Stages}

We will prove two main results here:  First, we will prove that in
order to comply with SFBC a Type 1 stage must always return a
result, \ie\ there can be no cases where a Type 1 method fails to
return a result because the conditions checked indicate that $c_1$
beats $c_2$ ($(\p, \N_{12})>0$), $c_2$ beats $c_3$ ($(\p,
\N_{23})>0$), and $c_3$ beats $c_1$ ($(\p, \N_{31})>0$).  Such
failures to return a result are related to the familiar cyclic
paradox described by Condorcet\cite{saari1995}.  Second, we will
prove that methods of that sort are point systems, in which
candidates are assigned points according to the ranks given by
voters, and the candidate with the most points wins.

\subsection{Type 1 Stages and Paradoxes}

We will begin by supposing that some stage of an election method,
numbered $s$ in the sequence of stages, is a Type 1 stage, and we
will assume that this stage does not always return a result.  If the
vectors normal to the boundaries defined in stage $s$ are denoted
$\N_{ij}^s$, then we can find a cyclic region of the unit simplex
where $(\p,\N_{12}^s)<0$, $(\p,\N_{23}^s)<0$, and
$(\p,\N_{31}^s)<0$.  In this case, we need to check the subsequent
stage $s+1$.  We will make no assumptions about whether stage $s+1$
is of Type 1 or some other type.  We will suppose, without loss of
generality, that stage $s+1$ selects candidate $c_3$ as the winner,
implying that $(\p,\N_{32}^{s+1})>0$ and $(\p,\N_{31}^{s+1})>0$.  As
long as one of the vectors $\N_{12}^{s}$, $\N_{23}^{s}$, and
$\N_{31}^{s}$ cannot be written as a linear combination of the other
two (a condition assured by our assumption that $(\p,\N_{12}^s)<0$,
$(\p,\N_{23}^s)<0$, and $(\p,\N_{31}^s)<0$), it is possible to find
a point where the boundaries defined in stage $s$ intersect.  Our
neutrality criterion then assures that it is possible for this
intersection to coincide with a region of profile space where stage
$s+1$ selects candidate $c_3$.

Now consider the hypersurface consisting of the intersection of the
boundaries defined by the normal vectors  $\N_{12}^{s}$,
$\N_{23}^{s}$, and $\N_{31}^{s}$.  We specifically consider a
portion of this hypersurface lying in a region where stage $s+1$
selects candidate $c_3$. (Below we will address the issue of whether
this intersection hypersurface lies inside the unit simplex.) Assume
initially that the profile vector is on the hypersurface defined by
the intersection of those boundaries. We can then displace the
profile vector slightly away from that intersection, so that we are
in a region where stage $s$ selects candidate $c_1$. Subsequently,
we can change $\p$ slightly so that it crosses the $1-2$ boundary
defined in stage $s$ (\textit{without crossing the $2-3$ boundary!})
and we now have a situation in which stage $s$ yields no winner and
so $c_3$ wins. However, this is a violation of SFBC, because the
outcome changed from $c_1$ to $c_3$ by crossing a boundary that does
not satisfy necessary conditions for the $1-3$ boundary in
SFBC-compliant methods.

In order to prove that this is a violation of SFBC, however, we must
address whether this intersection of planes occurred within the unit
simplex (\ie\ did it occur for a valid profile?).  A paradox that
only occurs when the profile vector $\p$ is outside the unit simplex
(\eg\ some components of $\p$ are negative) is of no consequence for
election methods, since we cannot have a negative number of voters
submitting a ballot of a particular type.  We will address this by
considering what happens when the normal vectors are changed in a
way that shifts the intersection out of the unit simplex.

Suppose initially that the boundaries all meet in the center of the
simplex at the point $\I/d$ discussed above.  Our analysis above, in
which we displace the profile slightly from the intersection,
clearly applies.  Suppose that we then start shifting boundaries
around by changing the right-hand side of each condition, so that
the victory condition is now $(\p,\N^s_{ij})>\delta^s_{ij}$.  This
will move each boundary around in a direction parallel to its normal
vector.  However, we have seen previously that changing the normal
vector another vector proportional to $\I$.  Adding to a normal some
vector proportional to $\I$ does not affect compliance with SFBC,
because all of the elements of a normal vector change by the same
amount, so our conditions regarding maximum and minimum elements are
still met. We also change the profile so $\p$ so that its position
relative to the intersection is unchanged.

Even if we can translate the cyclic region entirely outside the unit
simplex, however, there is a second cyclic region, that intersects
the original cyclic region only along a surface of dimension 2 less
than the dimension of the boundary.  This second cyclic region
cannot be reached by crossing only a single boundary, as it involves
a different intransitive relationship, one in which each pairwise
comparison has been reversed.  Translating the first cyclic region
outside the simplex does not translate the second cyclic region
outside, and because the cyclic regions extend infinitely far out
from their intersection point, it is impossible to translate both
regions outside of the simplex.

It therefore follows that SFBC violations are inevitable if a
condition $(\p,\N^s_{ij})<0$ (note the ''less than" symbol in the
inequality) causes a stage to return no winner in a case where none
of the previous stages of the method returned a winner, and the
normal vector $\N_{ij}$ only meets the requirements for a single
type of boundary in an SFBC-compliant method, \ie\ it is a Category
1 normal vector.  This gives us the following theorem:

\begin{theorem}\label{thm:type1a}
In a voting method that complies with SFBC, if one of the normal
vectors defining an inequality only satisfies the requirements for a
single boundary, non-satisfaction of that inequality cannot cause
the stage to return no winner (unless there is a tie and none of the
subsequent stages return a winner).
\end{theorem}

This may seem like a surprising result, because if a stage contains
an inequality then there should be cases where that inequality is
not satisfied; otherwise there would be no reason to include that
inequality.  However, the non-satisfaction of an inequality need not
cause the entire stage to return no winner.  It could be that
non-satisfaction of a particular inequality just means that the
stage returns some other winner.  Suppose we take a particular
victory condition in a stage and examine the vectors defining the
associated inequalities.  Normal vectors can be sorted into the
three categories (enumerated at the end of Section
\ref{sec:vectors}, depending on whether they satisfy the
requirements for only a single boundary (category 1) while other
vectors may satisfy the requirements for multiple boundaries
(categories 2 and 3, depending on how many requirements are
satisfied).  Theorem \ref{thm:type1a} tells us that if we examine
all of the possible victory conditions for the various candidates in
a given stage, in at least one of those conditions \textbf{all} of
the inequalities involving vectors of the first category will be
satisfied.  Whether or not the stage returns a winner then depends
on the satisfaction of inequalities involving vectors of the second
and third categories.

One consequence of Theorem \ref{thm:type1a} is that if all of the
vectors defining a stage of conditions are of the first category
(\ie\ each vector only satisfies the requirements for a single
boundary in an SFBC-compliant method), then there can be no case
where the non-satisfaction of any of the inequalities in that stage
causes the stage to return no winner.  This leads to the following
Theorem:

\begin{theorem}\label{thm:type1b}
If an SFBC-compliant election method includes a Type 1 stage, that
stage must be the last stage of the method.
\end{theorem}

\subsection{Type 1 stages must return winners}

Proving that a Type 1 stage of an election method must be the final
stage does not address the issue of whether a type 1 stage must have
a winner.  One could consider the possibility of an election method
in which the final stage is a Type 1 stage that only gives paradoxes
(no winner) in situations where some previous stage has already
yielded a winner.  We will now prove that such paradoxes lead to
violations of the SFBC, because constructing a method to avoid such
paradoxes imposes requirements on the earlier stages that are
incompatible with the SFBC.

Suppose that stage $s$ of an SFBC-compliant method is of Type 1.
Theorem \ref{thm:type1b} says that it must be the last stage of the
election method.  Stage $s$ is only utilized if all of the previous
stages returned no winner.  A necessary condition for a previous
stage to return no winner is for a number of inequalities (involving
vectors of the second or third categories defined above) to be
unsatisfied.  We could therefore generate a number of new stages
that involve all of the conditions and inequalities in $s$,
augmented with inequalities indicating that other stages returned no
winners.  Those inequalities would involve vectors of the second and
third categories, so these new new stages will be Type 1b or Type 3.

There would be a great number of possible ways to reach stage $s$
(depending on which inequalities were unsatisfied in the previous
stages) so there would be a great many new stages formed by
augmenting $s$.  However, these new stages could be inserted
\textit{anywhere} in our specification of the procedure without
changing the election result.  We could thus place one of these
augmented stages at the beginning of the procedure.

If there are cases where stage $s$ has a paradox like that of
Condorcet (\ie\ $(\p, \N_{12})<0$, $(\p, \N_{23})<0$, and $(\p,
\N_{31})<0$) then there will be cases where the first stage of the
election method fails to return a winner because of an unsatisfied
inequality involving a vector of the first category.  According to
Theorem \ref{thm:type1b}, this leads to a violation of SFBC.  This
leads to the following Theorem:
\begin{theorem}\label{thm:type1c}
A Type 1 stage in an election method satisfying SFBC must always
return a winner (except in the case of ties).  There can be no
Condorcet-type paradoxes in a Type 1 stage of an election method
satisfying SFBC.
\end{theorem}

Finally, we can show that the SFBC imposes an additional constraint
on Type 1 stages.  One could suppose that a Type 1 stage $s$
sometimes returns multiple winners (\ie\ conditions for different
candidates to win are not mutually exclusive) but that these
multiple victories only happen in situations where a previous stage
has returned a winner, avoiding the difficulty of multiple winners.
We will use reasoning analogous to that of Theorem \ref{thm:type1c}
to show that a Type 1 stage must always return a unique winner
(except in the case of ties, which are a lower-dimensional subset of
the unit simplex).

Suppose that we generate new stages by augmenting the conditions of
stage $s$ with linear inequalities that are satisfied when other
stages return no winners.  The inequalities added to the augmented
stages will again involve vectors of the second or third categories,
so the augmented stages are Type 1b or Type 3.  There will again be
a great number of augmented stages, reflecting the different ways
that stages prior to $s$ could fail to return a winner.  However, we
must also augment the conditions of stage $s$ with additional
inequalities specifying that none of the other candidates win in
stage $s$.  These inequalities can be obtained by taking
inequalities from other conditions of stage $s$ and multiplying the
vectors by $-1$.  The non-satisfaction of one of these inequalities
(involving normal vectors of Category 1) will cause the stage to
return no winner, in violation of Theorem \ref{thm:type1a}.  This
gives our final theorem of this section:
\begin{theorem}
A stage of Type 1 in an election method satisfying SFBC must always
return a unique winner, except in the case of ties.
\end{theorem}

\subsection{Generators of Type 1 Stages must not produce paradoxes}

We will prove this result by constructing stages that have multiple
generators, some of them with paradoxes and showing that they lead
to problems that can only be solved by introducing an additional
single-generator rule that decides the outcome in all cases.  A
generator defines a series of relationships (possibly intransitive)
between candidates of the form $c_i$ beats $c_j$.  Suppose that a
method has at least 2 generators, and we call 2 of these
$\mathbf{A}$ and $\mathbf{B}$. We have 3 possibilities: Both
generators always generate transitive relations, both can generate
intransitive relations, and one always generates a transitive
relation while the other can generate an intransitive relation.
Additionally, we consider the possibility that the stage has more
than one condition for each candidate to win, as well as the
possibility that the stage has exactly one condition for each
candidate to win.

Now consider stages in which each condition has inequalities
produced from different generators.  Suppose that all of the
conditions for $c_1$, $c_2$, and $c_3$ to win have inequalities of
the form $(\p,\N_{12}^A)$,  $(\p,\N_{23}^A)$, and $(\p,\N_{31}^A)$,
and each is supplemented by an inequality generated from
$\mathbf{B}$.  (These inequalities might also be supplemented by
inequalities from other generators, but in what follows we get
sufficient conditions on $\mathbf{A}$ and $\mathbf{B}$.)  If
$\mathbf{A}$ sometimes gives a paradox then this stage can return no
winner.  It therefore follows that $\mathbf{A}$ must always return a
winner, and the only candidate who can win this stage is the
candidate elected according to $\mathbf{A}$.  The inequalities
generated by $\mathbf{B}$ are either superfluous or else lead to
paradoxes.

We are left with the possibility of producing a Type 1 stage by
combining inequalities from different generators in each condition.
Considering only swaps of candidates $c_1$, $c_2$, and $c_3$, we get
6 possible conditions from $\mathbf{A}$ and $\mathbf{B}$.  We list
the associated sets of inequalities below:

\begin{tabular}{|c|c|c|}
    \hline Condition 1.1 & Condition 2.1 & Condition 3.1 \\
    $(\p,\N_{12}^A)>0$ & $(\p,\N_{23}^A)>0$ & $(\p,\N_{31}^A)>0$ \\
    $(\p,\N_{13}^B)>0$ & $(\p,\N_{21}^B)>0$ & $(\p,\N_{32}^B)>0$ \\ \hline
    Condition 1.2 & Condition 2.2 & Condition 3.2 \\
    $(\p,\N_{12}^B)>0$ & $(\p,\N_{23}^B)>0$ & $(\p,\N_{31}^B)>0$ \\
    $(\p,\N_{13}^A)>0$ & $(\p,\N_{21}^A)>0$ & $(\p,\N_{32}^A)>0$ \\ \hline
\end{tabular}
\\
\\
Suppose, without loss of generality, that $\mathbf{A}$ generates a
transitive relationship $c_1\succ c_2 \succ c_3$ and $\mathbf{B}$
generates an intransitive relationship $c_1\succ c_2$, $c_2\succ
c_3$, and $c_3\succ c_1$.\footnote{We are using the notation $c_i
\succ c_j$ and $c_j \prec c_i$ to mean ``$c_i$ is preferred to
$c_j$".  The notation $c_i \sim c_j$ will mean that neither
candidate is preferred to the other.} The stage described above
gives:

\begin{tabular}{|c|c|c|}
    \hline
    Condition 1.1 & Condition 2.1 & Condition 3.1 \\
    $c_1\succ c_2$ & $c_2\succ c_3$ & $c_3\prec c_1$ \\
    $c_1\prec c_3$ & $c_2\prec c_1$ & $c_3\prec c_2$ \\ \hline
    Condition 1.2 & Condition 2.2 & Condition 3.2 \\
    $c_1\succ c_2$ & $c_2\succ c_3$ & $c_3\succ c_1$ \\
    $c_1\succ c_3$ & $c_2\prec c_1$ & $c_3\prec c_2$ \\ \hline
\end{tabular}
\\
\\
In this case, $c_1$ satisfies Condition 1.2 and the inequalities
generated by $\mathbf{B}$ are thus superfluous.  If the cycle
generated by $\mathbf{B}$ were reversed, $c_1$ would still win, but
by Condition 1.l.

Finally, suppose that $\mathbf{A}$ and $\mathbf{B}$ give opposite
paradoxes with all relationships reversed.  We get the following
table of outcomes:

\begin{tabular}{|c|c|c|}
    \hline
    Condition 1.1 & Condition 2.1 & Condition 3.1 \\
    $c_1\succ c_2$ & $c_2\succ c_3$ & $c_3\succ c_1$ \\
    $c_1\succ c_3$ & $c_2\succ c_1$ & $c_3\succ c_2$ \\ \hline
    Condition 1.2 & Condition 2.2 & Condition 3.2 \\
    $c_1\prec c_2$ & $c_2\prec c_3$ & $c_3\prec c_1$ \\
    $c_1\prec c_3$ & $c_2\prec c_1$ & $c_3\prec c_2$ \\ \hline
\end{tabular}
\\
\\
The only way to decide this contest is to introduce an additional
generator $\mathbf{C}$ and augment the conditions with more
inequalities.  The outcome is then decided by $\mathbf{C}$ in this
case, and in order to avoid paradoxes, $\mathbf{C}$ must also return
the same winner as this stage would return without $\mathbf{C}$, in
cases where $\mathbf{A}$ and $\mathbf{B}$ together return a unique
winner.  It therefore follows that in all cases this stage must
return the result that $\mathbf{C}$ would return, and so
''$\mathbf{A}$ and $\mathbf{B}$ are superfluous.

We are therefore left to conclude that when constructing a Type 1
stage with multiple generators, none of the generators can give a
paradox.

\subsection{Type 1 Stages are Point Systems}

Given that a Type 1 stage must always return a unique winner (except
in the case of ties) and hence must always be the last stage in an
election method satisfying SFBC, and that the generators cannot give
rise to paradoxes, we now ask what sorts of election rules are
defined by generators of Type 1 stages. We will prove below that
such election rules are equivalent to point systems, in which a
candidate receives points based on the position assigned on a
ballot, irrespective of how other candidates are ranked on that
ballot, a candidate's points from all of the ballots are summed, and
the candidate with the most points wins. When the election method
requires voters to submit a complete ranking of candidates, with all
ranks used and no equal rankings, these methods are commonly called
''positional methods" \cite{saari1995}.  One obvious SFBC-compliant
election method is anti-plurality voting. Any other positional
methods that assigns equal (and maximum) points to first and second
choice candidates on a ballot would also comply with SFBC, as there
is no disincentive for a voter to list his or her sincere favorite
in first place:  If there is a close race between two candidates,
and neither of those candidates is a voter's favorite, that voter
can list his or her sincere favorite in first place and the more
preferred of the front-runners in second place, because the second
place candidate on the ballot will receive the maximum point total.

However, positional methods are not the only possible Type 1 stages.
In Section \ref{sec:ballotdesign}, we included in our analysis
ballots in which some positions may be left empty and other
positions may be assigned to multiple candidates, and we remarked
that Approval Voting is one such method.  Approval Voting does not
satisfy SFBC, because there are times when a voter has an incentive
to approve multiple candidates, giving equal points to his or her
favorite as well as a compromise candidate.  However, a modified
form of Approval Voting, with 3 ranks on the ballot and equal points
for the first and second places \textit{would} satisfy our phrasing
of SFBC, albeit on a technicality.  Likewise, we could modify Range
Voting\cite{Venzke,smith}, an SFBC-compliant technique in which
voters assign points to candidates within some specified range (\eg\
0 to 5).  In a modified SFBC-compliant form of Range Voting, there
would be two top positions on the ballot, with equal points but one
would be recorded as a voter's true favorite.  This illustrates yet
again that while SFBC can be satisfied in principle, it is difficult
to design an SFBC-compliant method that makes a meaningful
distinction between first and second place.

The fact that a Type 1 stage $s$ must \textit{always} return a
winner means that the stage will never give a paradox, which in turn
implies that there is always a transitive relationship among the
candidates.  If $(\p,\N^s_{12})=(\p,\N^s_{23})=0$, \ie ties between
$c_1$ and $c_2$ as well as $c_2$ and $c_3$, then $(\p,\N^s_{13})=0$,
(\ie\ there is also a tie between $c_1$ and $c_3$.  More generally,
if $(\p,\N^s_{12})=(\p,\N^s_{23})=...=(\p,\N^s_{n_{c}-1,n_c})=0$
implies that all other inner products $(\p,\N^s_{ij})$ are also
zero, then the $n_c(n_c-1)/2$ normal vectors $\{ N^s_{ij} \}$ are
not linearly independent.  This implies that of all the normal
vectors to the boundaries, \textit{at most} $n_c-1$ of them are
linearly independent.  We can show that no fewer than $n_c-1$ normal
vectors are linearly independent by considering ties between $c_1$
and $c_2$, $c_2$ and $c_3$, \etc\, all the way to $c_{n_c-2}$ and
$c_{n_c-1}$.  These conditions will mean that many pairs of
candidates are tied with each other (\eg\ $c_1$ and $c_{n_c-1}$ are
tied) but they do not guarantee a tie between $c_{n_c-1}$ and
$c_{n_c}$.

Due to the Neutrality criterion, in a stage with exactly 1 victory
condition per candidate and $n_c-1$ inequalities per candidate, we
also have that each boundary normal $\N^s_{ij}$ must be unchanged by
swaps of any candidates other than $c_i$ and $c_j$.  It therefore
follows that $(\p,\N^s_{ij})$ depends only on how many voters
list$c_i$ and $c_j$ in each spot on the ballot.  We could therefore
write $(\p,\N^s_{ij})=T_i-T_j$ where $T_i$ and $T_j$ are point
totals assigned to each candidate according to how many voters list
that candidate in a particular spot on the ballot.  There are $n_c$
point totals to calculate, and if we calculate the differences
between all possible pairs of point totals we get $n_c-1$ linearly
independent quantities, which correspond to the $n_c-1$ linearly
independent normal vectors that can be found.

Moreover, if a point system is to satisfy SFBC, it must award equal
points to a voter's first and second choices, so that the voter can
give the maximum possible points to a contender in a close race
without having to demote his/her sincere favorite from the top place
on the ballot.  However, as we have discussed above, this is only a
technical case of compliance with SFBC, since the method makes no
practical distinction between first and second place.  Hence, Type 1
stages only satisfy ``Strong" FBC in a very weak sense.

The only remaining question to ask is whether we could use two or
more different Type 1 generators, each giving rise to a different
point system, to produce a Type 1 stage with multiple conditions for
a candidate to win.  However, in that case, a condition would be a
mix of inequalities related to different point systems.  In general,
there will be cases in which candidates $c_1$ and $c_2$ both beat
$c_3$ in two different point systems generated by vectors
$\mathbf{A}$ and $\mathbf{B}$, but $\mathbf{A}$ selects $c_1$ over
$c_2$, while $\mathbf{B}$ selects $c_2$ over $c_1$. Using the same
table of outcomes as above, we get:

\begin{tabular}{|c|c|c|}
    \hline
    Condition 1.1 & Condition 2.1 & Condition 3.1 \\
    $c_1\succ c_2$ & $c_2\succ c_3$ & $c_3\prec c_1$ \\
    $c_1\succ c_3$ & $c_2\succ c_1$ & $c_3\prec c_2$ \\ \hline
    Condition 1.2 & Condition 2.2 & Condition 3.2 \\
    $c_1\prec c_2$ & $c_2\succ c_3$ & $c_3\prec c_1$ \\
    $c_1\succ c_3$ & $c_2\prec c_1$ & $c_3\prec c_2$ \\ \hline
\end{tabular}
\\
\\
In this case, a third generator is needed to generate another point
system to decide between $c_1$ and $c_2$, and this third generator
must, as before, always give the same results as the method
generated by $\mathbf{A}$ and $\mathbf{B}$, rendering those two
generators redundant with the third generator.  We can thus conclude
that a Type 1 stage can only have a single generator, which
generates a point system.

We therefore get this theorem, which is our primary result in this
paper:

\begin{theorem}
    Any Type 1 stage that satisfies SFBC, Neutrality,
    Anonymity, Linearity, No Turnout Quota, and Decisiveness is a
    point system in which candidates receive points based on how
    many voters list the candidate in each spot on the ballot and the
    candidate with the most points wins.  In these point systems,
    the candidates listed in the first and second places on the
    ballot must receive equal points.
\end{theorem}
If the method requires a strict ranking (\ie\ no ties and no unused
ballot spots) of all $n_c$ candidates, then the method is a
Positional Method of the sort studied by Saari.

\section{Type 1b Stages}

%\textbf{Begin notes to self}
%
%There is a much simpler way to do this:
%
%If A satisfies $(\p,\N_{ax})>0$ and B satisfies $(\p,\N_{ax})>0$
%then a Type 1 vector of type Nab decides between them.
%
%Could we do a method where we do a pairwise comparison between 2
%candidates who exceed a threshold of points?  Becomes complicated in
%4-candidate case (see below), but for now consider 3 candidates.  No
%way for all 3 candidates to have votes from more than $2/3$ of
%voters.  Do a pairwise comparison between them.  Otherwise, use
%anti-plurality.
%
%Problem:  Say that A and B have the most points.  You prefer A to B
%and C to B.  B wins this stage, but C would win if it went to next
%stage.  You have an incentive to take points away form A so that it
%goes to next stage.
%
%\textbf{End notes to self}

We now consider Type 1b stages. Although a stage of an
SFBC-compliant method can never fail to return a winner due to the
non-satisfaction of an inequality specified by a Type 1 vector, a
Type 1 stage (based on a point system) could be augmented by
additional inequalities if these inequalities involve Type 2
vectors.  Because the underlying Type 1 stage would always return a
winner, the only way the stage would fail to return a winner is
through the non-satisfaction of an inequality specified by a Type 2
vector.

However, we must be careful here:  Suppose that candidate $c_1$
satisfies all of the Type 2 inequalities required for victory, and
all but one of the Type 1 inequalities required for victory. The
remaining inequality is not satisfied because the profile is on a
boundary:  $(\p,\N^s_{12})=0$.  The transitivity of the
relationships defined by the Type 1 inequalities implies that $c_2$
also satisfies all of the Type 1 inequalities required for victory
in stage $s$, except that $(\p,\N^s_{21})=0$.

In this case, changing the profile to change the sign of
$(\p,\N^s_{12})$ should change the outcome from $c_1$ to $c_2$
without requiring the use of a subsequent stage of conditions. This
only works if, whenever the profile is on the $1-2$ boundary defined
by the Type 1 inequalities (\ie\ $c_1$ and $c_2$ have equal points),
\textit{and} $c_1$ satisfies a Type 2 inequality, $c_2$ also
satisfies the analogous Type 2 inequality.  Because the Type 1
inequalities are expressed in terms of point totals, the Type 2
inequalities must then also be expressed in terms of point totals.
In other words, when $c_1$ and $c_2$ have equal points, and have
more points than any other candidate, both candidates must either
satisfy the Type 2 inequalities or both candidates must not satisfy
the Type 2 inequalities.

It then follows that the Type 2 inequalities can only depend on the
total number of points that a candidate receives, rather than a
comparison of the points received by 2 different candidates.  In
order to satisfy the No Turnout Quota criterion, the criterion must
be a threshold of support points proportional to the number of
voters participating, rather than some fixed number of points
independent of the number of voters. We then get the following
result:

\begin{theorem}
    Any Type 1b stage that satisfies SFBC, Neutrality,
    Anonymity, Linearity, No Turnout Quota, and Decisiveness is a
    point system in which candidates receive points based on how
    many voters list the candidate in each spot on the ballot and the
    candidate with the most points wins \textit{if the number of points
    received by that candidate exceeds a threshold that is
    proportional to the number of voters}.  In these point systems,
    the candidates listed in the first and second places on the
    ballot must receive equal points.
\end{theorem}
Consider an example of a Type 1b voting rule that satisfies SFBC:
Voters rank candidates, and a candidate receives 1 point from each
voter who ranks that candidate in first \textit{or} second place. If
the points received exceed a quota (\eg\ at least $75\%$ of the
voters give that candidate a point).  Otherwise, the winner is the
candidate ranked in last place by the \textit{fewest} voters.

Another example would be a method in which ballots have 4 places,
and the option to list no candidates in some of the places.  The
winner is the candidate ranked in first or second place by the
greatest number of voters, if that candidate is ranked in those
places by a majority of the voters.  Otherwise, the candidate ranked
in first, second, or third place by the greatest number of voters is
the winner.  While this method does not make a meaningful
distinction between first and second place, it does not obligate the
voter to put a candidate in second place.  There will be situations
in which a voter has a strategic incentive to rank a candidate in
second place (usually because his or her favorite is not a
contender, but some other less-preferred candidates are contenders),
but there will also be situations in which a voter has no such
incentive (usually because his or her favorite \textit{is} a
contender).  In the later case, the voter is able to make a
meaningful distinction between first place and the next most
highly-ranked candidate on the ballot.

%Interestingly, if we have 3 candidates and we only allow strict
%rankings (\ie\ 2 candidates cannot be ranked equally) Type 1b stages
%are not usable.  To see why, first note that only 1 type of
%SFBC-compliant positional method is possible with 3 candidates, and
%that method is ``top 2 voting", in which the voter's favorite and
%second favorite each get 1 point.  This is equivalent to
%anti-plurality voting for 3 candidates, in which the candidate
%ranked last by the fewest voters wins.  To construct a Type 1b stage
%from this method, we would augment it with a threshold requirement.
%However, if the threshold requirement is to be meaningful it must be
%greater than $2/3$ of the number of voters. To see this, note that
%in anti-plurality voting the candidate ranked last by the fewest
%voters will be ranked last by less than $1/3$ of the voters, which
%means that the candidate will be ranked first or second by at least
%$2/3$ of the voters.  Any threshold below $2/3$ of the number of
%voters will thus have no practical significance.  However, if the
%threshold is high enough to be meaningful, then sometimes that Type
%1b stage will return no winner, necessitating another stage to
%follow it.

\section{Type 2 Stages}

General results are harder to find for Type 2 stages, in which
\textit{all} of the inequalities are defined by Type 2 vectors, \ie\
the vectors defining the conditions for candidate $c_1$ to win
satisfy the conditions for the normals to the $1-2$, $2-3$, \etc\
boundaries.  However, we can show that if we restrict our attention
to 3-candidate elections in which voters must submit strict and
complete rankings (\ie\ all candidates are ranked and no two
candidates are ranked equal to each other) then such methods must be
non-monotonic. We will use the following definition of monotonicity,
taken from Saari \cite{saari1995}:

\begin{definition}
A voting method is \textbf{monotonic} if when $c_j$ is chosen with
some profile $\p$, and the only voters to change preferences change
them to give $c_j$ a higher ranking (while preserving the relative
rankings of all other candidates) then $c_j$ is still elected with
the new profile $\p'$.
\end{definition}

Let us examine the implications of this requirement for a normal
vector $\mathbf{v}$ that satisfies the requirements of SFBC for
$\N_{1,2}$ and $\N_{1,3}$ (\ie\ $\mathbf{v_1}$ is a Type 2 vector).
Suppose that candidate $c_3$ is elected and a voter has submitted a
ballot with the preference $c_1\succ c_2\succ c_3$.  If that voter
changes his ballot to read $c_1\succ c_3\succ c_2$ then $c_3$ should
still win rather than $c_1$. This implies that the component of
$\mathbf{v_1}$ corresponding to $c_1\succ c_2\succ c_3$ is greater
than or equal to the component corresponding to $c_1\succ c_3\succ
c_2$. However, this same vector also satisfies the conditions for
the $1-2$ boundary.  Suppose that we now have a situation where
$c_2$ wins (under the same conditions, \ie\ the outcome is
determined by the sign of the inner product $(\p,\mathbf{v_1})$) and
a voter then changes his ballot from reading $c_1\succ c_3\succ c_2$
to $c_1\succ c_2\succ c_3$.  In this case, $c_2$ should still win
rather than $c_1$, and so the component of $\mathbf{v_1}$
corresponding to $c_1\succ c_3\succ c_2$ should be greater than or
equal to the component corresponding to $c_1\succ c_3\succ c_2$.  We
therefore conclude that the components corresponding to $c_1\succ
c_2\succ c_3$ and $c_1\succ c_3\succ c_2$ must be equal.

SFBC also implies that these components must be the largest
components of the vector, so that voters whose favorite is $c_1$ do
not have to list their favorite below first place in order to submit
a ballot that gives $c_1$ the maximum benefit.  For simplicity, we
will assume that these components are both $+1$.  Furthermore, SFBC
implies that 2 other components must also be equal to $+1$, with one
of those components corresponding to a ballot that lists $c_2$ in
first place, and the other corresponding to a ballot that lists
$c_3$ in first place.  We will consider the implications of
monotonicity to determine which components should be $+1$.

Suppose that $c_1$ wins in a situation where the outcome is
determined by the sign of the inner product $(\p,\mathbf{v_1})$.  If
a voter changes his or her ballot from $c_2\succ c_3\succ c_1$ to
$c_2\succ c_1\succ c_3$, the candidate $c_1$ should still win,
implying that the component corresponding to $c_2\succ c_1\succ c_3$
must be larger than the component corresponding to $c_2\succ
c_3\succ c_1$, and hence must be equal to $+1$.  A similar analysis
implies that the component corresponding to $c_3\succ c_1\succ c_2$
must also be $+1$.  The other components, corresponding to $c_2\succ
c_3\succ c_1$ and $c_3\succ c_2\succ c_1$ must be equal to some
negative number $-m$.  We hence see that in the case of
SFBC-compliant and monotonic election methods with only 3 candidates
and strict rankings, all of the Type 2 normal vectors are of the
form:
\begin{equation}
    \mathbf{v_1} = \left( +1, +1, +1, -m, -m, +1  \right)
\end{equation}
We are working in the basis defined by Saari \cite{saari1995}:

\begin{center}
\begin{table}
\caption{Basis for SFBC-compliant methods with 3 candidates}
\begin{tabular}{|c|c|} \hline
Preference & vector \\ \hline $c_1\succ c_2\succ c_3$ & $\left(1, 0,
0, 0, 0, 0 \right)$ \\ \hline $c_1\succ c_3\succ c_2$ & $\left(0, 1,
0, 0, 0, 0 \right)$
\\ \hline $c_3\succ c_1\succ c_2$ & $\left(0, 0, 1, 0, 0, 0 \right)$
\\ \hline $c_3\succ c_2\succ c_1$ & $\left(0, 0, 0, 1, 0, 0 \right)$ \\ \hline
$c_2\succ c_3\succ c_1$ & $\left(0, 0, 0, 0, 1, 0 \right)$ \\ \hline
$c_2\succ c_1\succ c_3$ & $\left(0, 0, 0, 0, 0, 1 \right)$ \\ \hline

\end{tabular}
\end{table}
\end{center}

A similar form can be obtained for a Type 2 vector $\mathbf{v_2}$ or
$\mathbf{v_3}$ that expresses a condition for $c_2$ or $c_3$ to win.
 This sort of condition could be expressed as ``$c_1$ wins if the
number of voters listing $c_1$ in first or second place is greater
than $m$ times the number of voters listing $c_1$ in last place." It
thus follows that if an SFBC-compliant method is monotonic, and if
there are only 3 candidates and the ballots require voters to give
complete rankings of the candidates without equal rankings (in any
position) then there is only one possible form for a Type 2
condition.

The problem is that in this case the Type 2 conditions for $c_1$ to
win and for $c_2$ to win are not mutually exclusive.  This is easy
to show by example.  Suppose that all of the voters list $c_3$ in
last place.  It then follows that the number of voters listing $c_1$
in first or second place is greater than $m$ times the number
listing $c_1$ in last place (since that number is zero), and the
same holds true for the number listing $c_2$ in first or second
place.  This Type 2 stage can therefore not decide between $c_1$ and
$c_2$, violating our decisiveness condition.  We therefore get the
following result:
\begin{theorem}
    If an SFBC-compliant method is monotonic and satisfies Anonymity,
    Neutrality, Linearity, Decisiveness, and No Turnout Quota, it
    cannot have a Type 2 stage if there are 3 candidates and the
    ballots require voters to rank all candidates without any equal
    rankings.
\end{theorem}

\section{Ties and SFBC}
So far, we have largely neglected ties except to note their
existence and equate them to boundaries between victory regions in
profile space.  However, we have not considered the rules used to
break ties.  Interestingly, in the case of tie-breaking procedures,
it is possible to satisfy Strong FBC rather than just a weak form.
Suppose that we have some SFBC-compliant method, and we supplement
it with a rule that when there is a 2-way tie (\ie\ two boundaries
intersect) the winner is whichever of the 2 candidates is ranked
above the other by the most voters.  There is no incentive to list
another candidate ahead of one's favorite in this tie-breaking
procedure, and there is no incentive to list another candidate ahead
of one's favorite in any other part of the voting procedure (due to
SFBC-compliance).  However, the tie-breaking procedure truly
satisfies SFBC in a strong sense.  Notably, pairwise comparisons
work in ties, but not in general multi-candidate elections, because
ties are 2-way elections, whereas a general multi-candidate election
is susceptible to the Condorect paradox.  Unfortunately, this
stronger form of compliance with SFBC only happens in an exceedingly
rare case.

\section{FBC \vs\ SFBC}

Interestingly, while we can show that election methods are highly
restricted if we insist that no voter \textit{ever} have an
incentive to rank another candidate equal to his or her sincere
favorite, a much wider variety of election methods are possible if
we relax SFBC to FBC, and consider methods in which a voter
\textit{sometimes} has an incentive to rank another candidate
\textit{equal} to his or her sincere favorite, but never has an
incentive to rank another candidate \textit{above} his or her
sincere favorite. Interestingly, some of these methods actually make
meaningful distinctions between first and second place, unlike
methods that satisfy our very strict formulation of SFBC as
discussed above.

We will illustrate this point with three examples:

\subsection{Range Voting}

In Range Voting\cite{Venzke,smith}, each voter assigns each
candidate points on some scale (typically 0 to some upper bound),
and the candidate with the most points wins.  In the case where the
upper bound is $1$, Range Voting is equivalent to Approval Voting.
A voter may have an incentive to assign the maximum score to some
candidate other than his or her sincere favorite (if that candidate
is in a close race with a less-preferred candidate) but there is
never a disincentive to give the top score to the sincere first
choice.  This is a point system, just like the Type 1 SFBC-compliant
methods, and is a Type 1 FBC-compliant method.

\subsection{Majority Choice Approval}

A very simple example of a Type 1b FBC-compliant method is Majority
Choice Approval Voting.  In this simple method, a voter can rate
each candidate as ``Preferred", ``Approved", or ``Disapproved." The
candidate who is rated ``Preferred" by the greatest number of voters
wins, provided that he or she is rated thus by a \textit{majority}
of voters.  Otherwise, the candidate with the greatest combined
``Preferred" and ``Approved" ratings wins.  The comparisons of vote
totals are all expressed by Type 1 vectors, and the requirement that
the total exceed a majority threshold is expressed by a Type 2
vector, giving a Type 1b method.

\subsection{Majority Defeat Disqualification Approval}

Another FBC-compliant method is Majority Defeat Disqualification
Approval (MDDA), first studied by Kevin Venzke\cite{Venzke,smith}.
In MDDA and related methods, voters rank as many candidates as they
wish, with equal rankings allowed; all unranked candidates are
treated as being ranked equal to each other in last place. All
ranked candidates are said to be approved. A candidate $c_i$ is said
to be dominated by a candidate $c_j$ if a majority of voters rank
$c_j$ above $c_i$. If one candidate dominates all other candidates,
that candidate wins. Because equal rankings are allowed, it is
possible that there will be no majority favoring $c_i$ over $c_j$
\textit{and} no majority favoring $c_j$ over $c_i$, in which case
neither candidate is dominated by the other.  If there are multiple
un-dominated candidates, or no un-dominated candidates, some other
method must be used.

In MDDA and related methods based on the concept of majority
dominance, the outcome is determined by whether or not a candidate
is dominated by another candidate.  Let us define a vector
$\mathbf{d}_{12}$ such that $(\p,\mathbf{d}_{12})>0$ if $c_1$ is not
dominated by $c_2$. The elements of the $\mathbf{d}_{12}$ are
summarized in Table 2.
\begin{centering}
    \begin{table}
        \caption{Ballot counts being compared to determine if $c_1$ dominates $c_2$.}
        \begin{tabular}{|c|c|} \hline
            Does not prefer $c_2$ to $c_1$ &  Prefers $c_2$ to $c_1$
            \\
            (vector component $ = +1$) & (vector component $=-1$) \\
            \hline
            $c_1\succ c_2\succ c_3$ & $c_2\succ c_1\succ c_3$ \\ \hline
            $c_1\succ c_3\succ c_2$ & $c_2\succ c_3\succ c_1$ \\ \hline
            $c_3\succ c_1\succ c_2$ & $c_3\succ c_2\succ c_1$ \\ \hline
            $c_1 \succ c_2 \sim c_3$ & $c_2 \succ c_1 \sim c_3$ \\ \hline
            $c_1 \sim c_3 \succ c_2$ & $c_2 \sim c_3 \succ c_1$ \\
            \hline
             $c_1 \sim c_2 \succ c_3$ & \\ \hline
             $c_3 \succ c_1 \sim c_2$ & \\ \hline
        \end{tabular}
    \end{table}
\end{centering}

For each candidate, there is at least one preference that
corresponds to a maximum (largest positive) element ($+1$) of
$\mathbf{d}_{12}$.  There are also minimum (largest negative)
elements corresponding to preferences that list $c_2$ and $c_3$ in
first place.  This is a type 2 vector.  For a stage in which the
outcome is determined only by comparing candidates and eliminating
dominated candidates, all of the vectors are Type 2, and we thus see
that a wider range of Type 2 methods are possible if we relax SFBC
to FBC.

If there is more than 1 undominated candidate, the outcome is
decided by a stage that elects the undominated candidate ranked last
by the fewest people.  In this case, Type 2 vectors (to determine
who is not dominated) are combined with Type 1 vectors (comparing
points based on last place rankings) to give a Type 1b stage.

\section{Least Favorite Promotion}

Interestingly, we can formulate an analogous criterion for voters
who are less interested in supporting their favorite and more
interested in avoiding the accidental election of their least
favorite due to strategic manipulations.  This fear may be
well-warranted in some cases:  In a 3-candidate election with an
SFBC-compliant method, promoting the least favorite to second place
may be a viable strategy for electing the voter's favorite over his
or her second choice, but such strategies always come at a risk.

If we were to pursue the same approach as used above for identifying
SFBC-compliant methods, and call the new criterion ``Least Favorite
Promotion", we would find that a major portion of the methods are
point systems of some sort or another.  Some types of point systems
(\eg\ Range Voting, Approval Voting, positional methods that give
equal points to first and second choices and zero points to last and
second-last choices) would satisfy both SFBC and also Least Favorite
Promotion.

\section{Conclusions}

In conclusion, we have shown that the Strong Favorite Betrayal
Criterion is exceedingly difficult to satisfy.  Two of the four
geometrical categories of methods are restricted to point systems.
The Type 2 methods have more variety, but the most interesting
methods are only found if SFBC is relaxed to allow for the
possibility of \textit{sometimes} ranking another candidate equal to
one's favorite, \eg\ Majority Defeat Disqualification Approval. When
SFBC is replaced with FBC, the resulting methods actually make
meaningful distinctions between first and second place, satisfying
the ``spirit" but not the ``letter" of SFBC.  The most significant
practical consequence of these results is that election reformers
who want to be free from disincentives against supporting their
sincere favorite above all others must accept systems in which
voters sometimes rank another candidate equal to their favorite
(either explicitly, in FBC, or implicitly in point systems that give
equal points to first and second place).

\textbf{Acknowledgements:}  I thank all of the members of the
Election Methods Mailing List for useful discussions of this
problem, especially Forest Simmons and Warren D. Smith.  I first
learned of the Favorite Betrayal Criterion from reading the website
ElectionMethods.org, where Mike Ossipoff and Russ Paielli noted that
it is satisfied by Approval Voting, but not by any other methods
that they had analyzed. (The site content, alas, has been changed
since then, and the articles on FBC can no longer be found there.)

%\bibliographystyle{ieee}
%\bibliography{sfbc}

\begin{thebibliography}{1}

\bibitem{gibbard1973}
A.~Gibbard,
\newblock ``Manipulation of voting schemes - general result'',
\newblock {\em Econometrica}, vol. 41, no. 4, pp. 587--601, 1973.

\bibitem{Satterthwaite1975}
M.~A. Satterthwaite,
\newblock ``Strategy-proffness and arrows conditions - existence and
  correspondence theorems for voting procedures and social welfare functions'',
\newblock {\em Journal of Economic Theory}, vol. 10, no. 2, pp. 187--217, 1975.

\bibitem{jackson2004}
Jesse~L. Jackson and James~D. Henderson,
\newblock ``Making elections better, and stopping divisiveness too'',
\newblock {\em The Boston Globe}, p. A23, 2004.

\bibitem{anderson2006}
Richard Anderson-Connolly,
\newblock ``Instant runoff voting can solve primary problem for counties'',
\newblock {\em Seattle Post-Intelligencer}, p.~B7, 2006.

\bibitem{Venzke}
\newblock ``Majority Defeat Disqualification Approval" in Electowiki,
\newblock
\begin{verbatim}http://wiki.electorama.com/wiki/Majority_Defeat_Disqualification_Approval.\end{verbatim}

\bibitem{smith}
Warren~D. Smith and Mike Ossipoff,
\newblock ``Survey of voting methods that avoid favorite betrayal'',
\newblock http://rangevoting.org/FBCsurvey.html.

\bibitem{brams1983}
Steven~J. Brams and Peter~C. Fishburn,
\newblock {\em Approval voting},
\newblock Birkhèauser, Boston, 1983.

\bibitem{saari1995}
D.~Saari,
\newblock {\em Basic geometry of voting},
\newblock Springer, Berlin ; New York, 1995.

\bibitem{Tideman2006}
Nicolaus Tideman,
\newblock {\em Collective decisions and voting the potential for public
  choice},
\newblock Ashgate, Aldershot, England; Burlington, VT, 2006.


\end{thebibliography}

\end{document}